\documentclass{iopjournal}

\usepackage{graphicx}
\usepackage{dcolumn}
\usepackage{bm}
\usepackage{color}
\usepackage[mathlines]{lineno}
\usepackage[dvipsnames]{xcolor}
\usepackage[utf8]{inputenc}
\usepackage[T1]{fontenc}
\usepackage{newtxtext,newtxmath}
\usepackage{multirow}
\usepackage{threeparttable}
\usepackage{booktabs}
\usepackage{float}
\usepackage{amsmath}
\usepackage{gensymb}
\usepackage{tabularx}
\newcolumntype{C}{>{\centering\arraybackslash}X}

\usepackage{cite}

\begin{document}

\title{Experimental Demonstration of a Balanced Travelling Wave Parametric Amplifier}

\author{Nikita Klimovich$^{1,*}$\orcid{0000-0002-5401-8260}, Peter K. Day$^2$\orcid{0000-0002-3670-9017}, Henry G. Leduc$^2$\orcid{0000-0002-6721-5271}, and Boon-Kok Tan$^{1}$\orcid{0000-0002-6252-9351}}

\affil{$^1$Department of Astrophysics, University of Oxford, Oxford, United Kingdom}

\affil{$^2$Jet Propulsion Laboratory, Pasadena, United States of America}

\affil{$^*$Author to whom any correspondence should be addressed.}

\email{nikita.klimovich@physics.ox.ac.uk}

\keywords{Travelling wave parametric amplifier, TWPA, kinetic inductance, balanced amplifier, image rejection}

\begin{abstract}
Superconducting travelling-wave parametric amplifiers (TWPAs) provide broadband gain, low added noise, and high dynamic range, making them attractive for applications requiring wide instantaneous bandwidth. This is crucial for amplifying multiplexed narrow band signals spread over a wide frequency range, such as qubit resonator arrays, or amplifying inherently wide band signals, such as in spectrometers or radiometers. In many cases, however, the useful bandwidth of a TWPA is limited by the production of the idler tone at an image frequency.  For three-wave-mixing (3WM) TWPAs, for example, an input signal at $f_s$ is indistinguishable from one at $f_p - f_s$, which can produce spectral ambiguity when the input spectrum is broadband or not known in advance. We experimentally demonstrate a balanced architecture that doubles the instantaneous image-rejected bandwidth available to a 3WM TWPA by spatially separating the amplified signal from the output pump and generated idler, which resolves the sideband degeneracy.  The architecture consists of two amplifiers connected between a pair of 180$^{\circ}$ hybrid couplers, with the signal and pump injected and output through the sum and difference ports, respectively. Using two kinetic-inductance TWPAs, we demonstrate approximately 20 dB of signal gain across nearly an octave of bandwidth, together with pump rejection below --20 dB and an average image rejection of approximately --15 dB over the operating band. These results establish balanced TWPAs as a practical approach to broadband parametric amplification, enabling improved contamination free bandwidth and eliminating the need for complex filtering.
\end{abstract}

\section{Introduction}

Superconducting travelling-wave parametric amplifiers (TWPAs) are attractive as first-stage cryogenic amplifiers because they combine broadband gain with near quantum-limited added noise and comparatively high dynamic range \cite{Petrovnin2024,Rodrigues2024,Frasca2024}. TWPAs exploiting either the nonlinear inductance of a superconducting transmission line or the current response of Josephson junctions have demonstrated broad gain bandwidths without the narrow-band resonant structure characteristic of many lumped-element parametric amplifiers \cite{Eom2012,Macklin2015,KowBell2026,Kaufman2023,ptb}. These properties are particularly beneficial for frequency-multiplexed detector systems, where a single amplification chain must process many channels simultaneously \cite{Beltran2025,Gao2021}. Kinetic-inductance TWPAs in particular have been of great interest for highly-multiplexed astronomical detector arrays due to their higher dynamic range \cite{howe} and are being used to improve the sensitivity of microwave kinetic-inductance detectors (MKIDs) by replacing the HEMT as the first stage in the amplifier chain\cite{Bockstiegel2014,Zobrist2019}. The low power dissipation of TWPAs compared to HEMT amplifiers could also enable larger arrays of both MKIDs and heterodyne receivers based on SIS and HEB mixoers.  More generally, extension of kinetic-inductance TWPAs towards millimetre and sub-millimetre wavelengths has been proposed as a route to lower-noise astronomical receiver systems \cite{Tan2024}.

A practical complication of phase-preserving parametric amplification is the generation of an idler. In a three-wave-mixing (3WM) amplifier, a signal at angular frequency $\omega_s$ and a strong pump at $\omega_p$ are coupled to an idler at $\omega_i$, with $\omega_p=\omega_s+\omega_i$ \cite{Kamal2009}. In a conventional TWPA, the amplified signal, residual pump, and generated idler all emerge from the same physical output. If the input spectrum is not known in advance, a feature at one frequency is accompanied by a correlated/mirrored image at the corresponding idler frequency, so the signal and idler bands are no longer distinguishable at the output. Signal/idler contamination is consequently one of the practical obstacles to deploying TWPAs in measurement systems outside tightly controlled laboratory readout chains \cite{Longden2024,Praquin2025}. Astronomical heterodyne spectroscopy provides a particularly clear example of an application that benefits from image rejection \cite{Mundia}.  The intermediate-frequency (IF) spectrum contains a weak and a priori unknown astronomical spectrum profile as well as continuum backgrounds. Ambiguity between a desired spectral component and its image thus directly complicates interpretation of the measured spectrum. For this reason, sideband-separating receiver architectures are widely used in millimetre-wave astronomy. The Atacama Large Millimeter Array (ALMA), for example, uses sideband-separating SIS mixers specifically to distinguish the IF output from the upper and lower sidebands while preserving broad IF coverage \cite{Kerr2014}. The upper and lower RF sidebands of the heterodyne mixers superimposed at the IF range are analogous to the signal and idler tones in a TWPA, hence the need to separate them into different output ports to avoid spectral confusion.

A TWPA that provides low-noise gain while routing its internally generated idler away from the amplified signal therefore offers an analogous benefit to sideband separating mixers. In the ideal case, separating the two parametric outputs removes signal--idler spectral confusion and allows the useful signal band to be treated independently of the generated image, rather than downstream filtering or sacrificing bandwidth to guard bands \cite{Brookes2021,Mercier2019}. The same property is also attractive for densely multiplexed cryogenic detector readout, where removing idler products from the signal output reduces spectral occupancy and simplifies the requirements placed on subsequent channelisation and filtering stages \cite{Perspective2021}. As the signal and idler are inherently entangled \cite{Greco2021,Vesterinen2023}, separating the two modes is also useful for two-mode squeezing measurements \cite{CastellanosBeltran2008, Esposito2022}, quantum computing \cite{Naaman2022,Lecocq2021,Abdo2013} and quantum communication protocols \cite{Samsonov2020}.

A balanced TWPA architecture offers a direct way to obtain this separation \cite{Nilsson2024}. The approach proposed in Ref.~\cite{Longden2024} places two nominally identical TWPAs in parallel between a pair of hybrid couplers and uses the relative phases imposed on the propagating tones to control the output port at which each component recombines. For the DC-biased 3WM configuration demonstrated here, the signal and pump are respectively injected through the sum and difference ports of the first $180^{\circ}$ hybrid. The signal is divided equally in phase between the two amplifier arms, whereas the pump is divided equally with a relative phase of $\pi$. Since the generated idler phase follows the difference between the pump and signal phases, the idlers generated in the two arms are also in anti-phase. A second $180^{\circ}$ hybrid then recombines the amplified signal constructively at one output port while directing the pump and idler to the other. In addition to image rejection, this arrangement provides a natural pump-cancellation mechanism, reducing the additional complication of pump filtering prior to the following cryogenic and room-temperature electronics \cite{Hougland2025}.

\section{Theory}
\label{sec:theory}

\begin{figure*}[!t]
    \centering
    \includegraphics[width=0.85\linewidth]{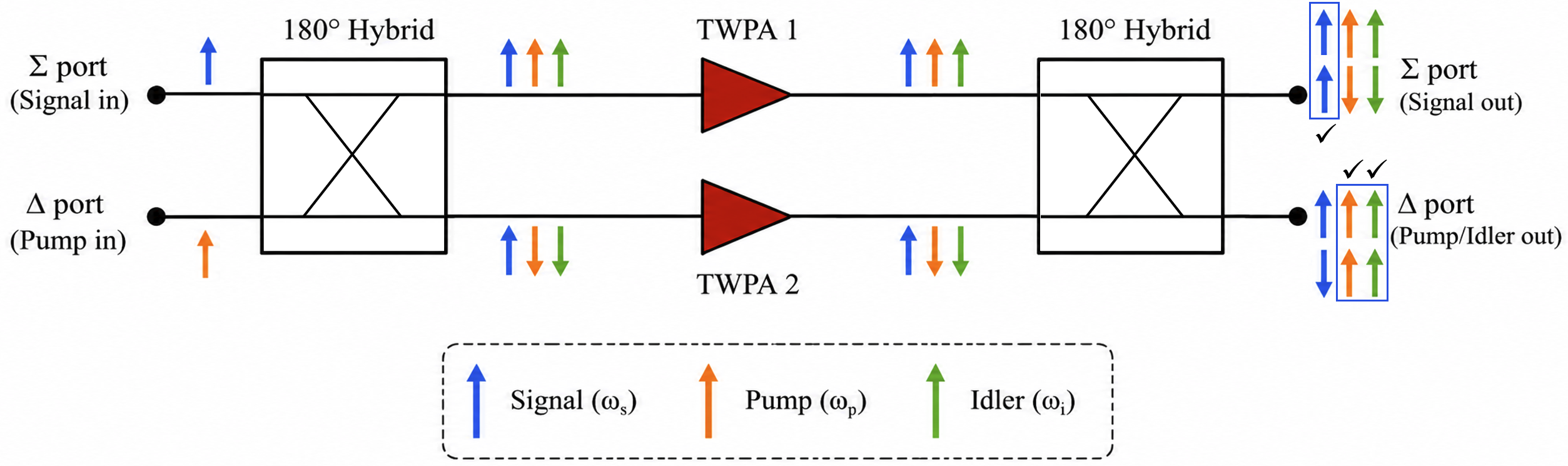}
    \caption{A schematic diagram showing a 3WM balanced TWPA configuration consisting of two identical TWPAs placed between two $180^\circ$ hybrids. Arrows indicate the phases of the signal, pump, and idler tones at various stages of the amplifier to demonstrate how the constructive and destructive interference in the two output ports results in signal, pump, and idler separation.}
    \label{fig:BalancedPlot}
\end{figure*}

A balanced TWPA consists of two identically designed TWPAs connected between a pair of hybrid couplers, as shown in Figure \ref{fig:BalancedPlot}. This configuration was proposed and described in detail in Ref.~\cite{Longden2024}. In this section, we briefly recreate the relations relevant to the three-wave-mixing (3WM) configuration. 

\subsection{Three-wave mixing}

For a 3WM TWPA, a strong pump at angular frequency $\omega_p$ amplifies an incident signal at $\omega_s$ while generating an idler at $\omega_i$, with energy conservation requiring
\begin{equation}
    \omega_p = \omega_s + \omega_i .
    \label{eq:3wm_energy}
\end{equation}

The current propagating along a TWPA may be expressed as a sum of slowly varying travelling-wave components,
\begin{equation}
    I(z,t) = \sum_{j=p,s,i} A_j(z) e^{-\gamma_j z} e^{-i(\omega_j t-\delta_j)} + \mathrm{c.c.},
    \label{eq:current_ansatz}
\end{equation}
where $A_j(z)$ is the slowly varying complex amplitude, $\gamma_j = \alpha_j+i\beta_j$ is the complex propagation constant and $\delta_j$ is the initial phase of each tone. Equivalently, the initial phase can be absorbed into the complex amplitude through
\begin{equation}
    A'_j(z) = A_j(z)e^{i\delta_j}.
\end{equation}

The phase of each incident wave must therefore be retained when considering the interference of the outputs from the two TWPAs. In the undepleted-pump approximation, the coupled-mode equations describing the signal and idler may be written in the generic form
\begin{align}
    \frac{dA_s}{dz}
        &= i\kappa_s A_p A_i^{*}
           e^{i\Delta k z} + \mathrm{c.c.},
           \label{eq:cme_signal}\\
    \frac{dA_i}{dz}
        &= i\kappa_i A_p A_s^{*}
           e^{i\Delta k z} + \mathrm{c.c.},
           \label{eq:cme_idler}
\end{align}
where $\kappa_s$ and $\kappa_i$ describe the nonlinear coupling strengths and $\Delta k = k_p - k_s - k_i$ is the phase mismatch from transmission line dispersion. For efficient amplification, $\Delta k$ is engineered to perfectly compensate for the pump self-phase modulation, but this is not strictly required for balanced operation.

From Eq.~(\ref{eq:cme_idler}), the idler amplitude is proportional to $A_p A_s^{*}$, and hence its phase obeys
\begin{equation}
    \delta_i = \delta_p-\delta_s + \delta_0 ,
    \label{eq:3wm_phase}
\end{equation}
where $\delta_0$ contains phase accumulated during propagation and any phase convention associated with the nonlinear coupling coefficient. Since $\delta_0$ is common to the two identical TWPAs, the relative idler phase between the two arms is
\begin{equation}
    \Delta\delta_i = \Delta\delta_p-\Delta\delta_s.
    \label{eq:3wm_relative_phase}
\end{equation}

\subsection{Balanced 3WM configuration}

The signal and pump tones are supplied to the $\Sigma$ and $\Delta$ ports of the first $180^\circ$ hybrid. For an ideal hybrid, its action on the two propagating modes can be represented, up to a common phase, by
\begin{equation}
    \begin{pmatrix}
        A_{1}\\
        A_{2}
    \end{pmatrix}
    =
    \frac{1}{\sqrt{2}}
    \begin{pmatrix}
        1 & 1\\
        1 & -1
    \end{pmatrix}
    \begin{pmatrix}
        A_{\Sigma}\\
        A_{\Delta}
    \end{pmatrix},
    \label{eq:180hybrid}
\end{equation}
where the subscripts 1 and 2 denote the output ports connected to the TWPA arms, and $\Sigma$ and $\Delta$ denote the corresponding in-phase and anti-phase hybrid ports.

The signal is injected into the $\Sigma$ port and is consequently divided equally and in phase between the two TWPAs,
\begin{equation}
    A_{s,1} = \frac{A_s}{\sqrt{2}},
    \qquad
    A_{s,2} = \frac{A_s}{\sqrt{2}},
    \label{eq:signal_split}
\end{equation}
such that
\begin{equation}
    \Delta\delta_s = 0.
\end{equation}

Conversely, the pump is injected into the $\Delta$ port, producing equal pump amplitudes with a relative phase of $\pi$,
\begin{equation}
    A_{p,1} = \frac{A_p}{\sqrt{2}},
    \qquad
    A_{p,2} = -\frac{A_p}{\sqrt{2}},
    \label{eq:pump_split}
\end{equation}
and therefore
\begin{equation}
    \Delta\delta_p=\pi.
\end{equation}

Applying Eq.~(\ref{eq:3wm_relative_phase}) then gives
\begin{equation}
    \Delta\delta_i
    =
    \Delta\delta_p-\Delta\delta_s
    =
    \pi.
    \label{eq:idler_phase_balanced}
\end{equation}
Thus, while the amplified signal fields remain in phase between the two TWPAs, both the pump and the generated idler are in anti-phase.

The second $180^\circ$ hybrid recombines the waves according to the same sum--difference relation. For an arbitrary pair of arm amplitudes $(A_1,A_2)$,
\begin{align}
    A_{\Sigma,\mathrm{out}}
        &= \frac{A_1+A_2}{\sqrt{2}},
        \label{eq:sum_out}\\
    A_{\Delta,\mathrm{out}}
        &= \frac{A_1-A_2}{\sqrt{2}}.
        \label{eq:diff_out}
\end{align}
Consequently, the in-phase signal components interfere constructively at the
$\Sigma$ output and destructively at the $\Delta$ output,
\begin{equation}
    A_{s,\Sigma} \neq 0,
    \qquad
    A_{s,\Delta}=0,
\end{equation}
whereas the anti-phase pump and idler components are routed to the opposite port,
\begin{equation}
    A_{p,\Sigma}=A_{i,\Sigma}=0,
    \qquad
    A_{p,\Delta},A_{i,\Delta}\neq0.
    \label{eq:frequency_separation}
\end{equation}
The balanced device therefore spectrally separates the outputs of the TWPA: the amplified signal exits one physical port while the pump and idler exit the other. For applications where the incident signal spectral profile is unknown (e.g., in heterodyne astronomical measurements), this acts analogously to a sideband separating mixer, effectively doubling the usable bandwidth of the TWPA by eliminating the signal/idler spectral confusion. 

For identical TWPAs with complex signal transmission $g_s$, the signal amplitudes immediately before the output hybrid are
\begin{equation}
    \mathbf{A}_s^{\mathrm{TWPA}}
    =
    \frac{g_s A_s}{\sqrt{2}}
    \begin{pmatrix}
        1\\1
    \end{pmatrix}.
\end{equation}
Application of Eq.~(\ref{eq:180hybrid}) gives
\begin{equation}
    \mathbf{A}_s^{\mathrm{out}}
    =
    \begin{pmatrix}
        g_s A_s\\0
    \end{pmatrix},
    \label{eq:signal_recombination}
\end{equation}
so that the gain measured at the signal port is, ideally, equal to the gain
of either constituent TWPA,
\begin{equation}
    G_s = |g_s|^2.
\end{equation}
This means that the two $3~\mathrm{dB}$ divisions experienced by the signal do not introduce an additional $3~\mathrm{dB}$ loss, since the amplified fields from the two arms recombine coherently at the signal output.

\subsection{Effect of differential phase and amplitude imbalance}
\label{sec:imbalanceTheory}

The previous section assumed that both TWPA arms and the two hybrids had equal amplitudes and the prescribed relative phases, resulting in perfect signal, pump, and idler separation. In a physical system, differences in TWPA electrical length, gain, cable length, hybrid amplitude balance, and hybrid phase balance produce finite leakage into the nominally isolated ports. These non-idealities can be represented by writing the two arm amplitudes for an anti-phase component as
\begin{equation}
    \mathbf{A}
    =
    \frac{A}{\sqrt{2}}
    \begin{pmatrix}
        1\\
        -r e^{i\epsilon}
    \end{pmatrix},
    \label{eq:imbalance_vector}
\end{equation}
where $r \neq 1$ describes the amplitude imbalance between the two arms and $\epsilon \neq 0$ is the error from the ideal $\pi$ relative phase. Recombination at the output hybrid gives
\begin{align}
    A_{\mathrm{leak}}
        &= A
           \left(1-r e^{i\epsilon}\right),\\
    A_{\mathrm{desired}}
        &= A
           \left(1+r e^{i\epsilon}\right).
\end{align}
The resulting power leakage relative to the desired output is therefore
\begin{equation}
    \frac{P_{\mathrm{leak}}}{P_{\mathrm{desired}}}
    =
    \frac{
        \left|1-r e^{i\epsilon}\right|^2
    }{
        \left|1+r e^{i\epsilon}\right|^2
    }.
    \label{eq:leakage_general}
\end{equation}
For small errors, $r=1+\eta$, $|\eta|\ll1$ and $|\epsilon|\ll1$, Eq.~(\ref{eq:leakage_general}) reduces approximately to
\begin{equation}
    \frac{P_{\mathrm{leak}}}{P_{\mathrm{desired}}}
    \simeq
    \frac{\eta^2+\epsilon^2}{4},
    \label{eq:leakage_small}
\end{equation}
and the same expression applies to the in-phase component after interchanging the desired and isolated output ports. This leakage power forms two of the three key metrics for evaluating the performance of a balanced TWPA:
\begin{enumerate}
    \item Signal gain, $G_s = |A_{s,\Sigma}^2(\text{out})/A_{s,\Sigma}^2(\text{in})|$,
    \item Pump rejection ratio, $R_p = P_{\mathrm{leak}}/P_{\mathrm{desired}}$ for the pump,
    \item Image rejection ratio, $R_i = P_{\mathrm{leak}}/P_{\mathrm{desired}}$ for the idler,
\end{enumerate}
In the following sections, we experimentally demonstrate the effectiveness of the 3WM balanced TWPA scheme evaluated using these criteria.  

\section{Measurements} 
\label{sec:measurements}

\subsection{Individual TWPAs}
\label{sec:device_design}

\begin{figure}[h]
    \centering
    \includegraphics[width=\linewidth]{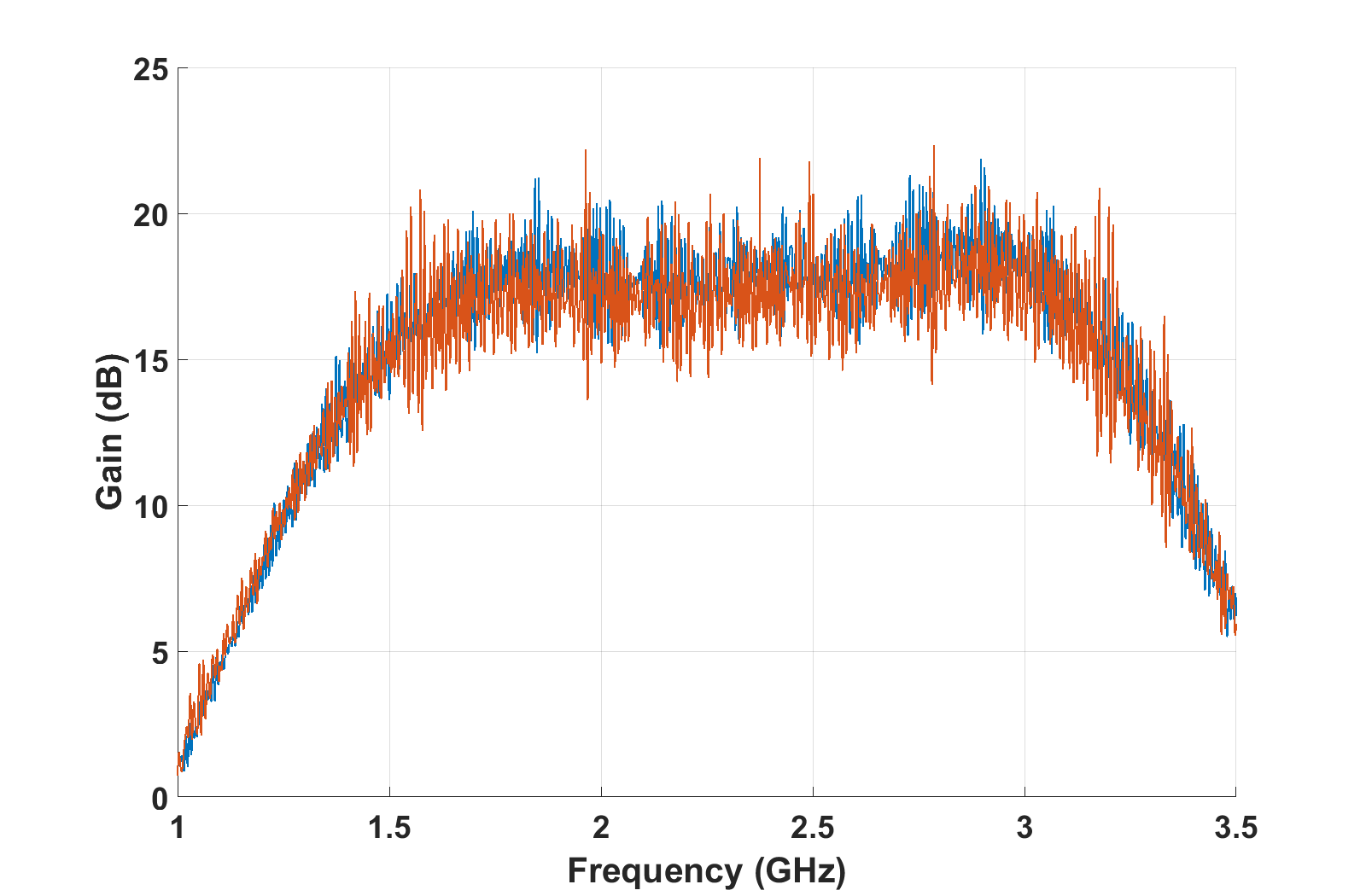}
    \caption{Individual gain profiles of the two TWPAs when supplied with an identical 4.75 GHz pump and 0.40 mA DC current.}
    \label{fig:IndividualTWPAGain}
\end{figure}

Our balanced amplifier is formed from two nominally identical Niobium Titanium Nitride (NbTiN) kinetic-inductance travelling-wave parametric amplifiers. They exhibit similar gain profiles when operated using identical pump and DC biasing conditions, as shown in Figure~\ref{fig:IndividualTWPAGain}. 

Further details on the design of these two TWPAs are provided in Appendix~\ref{app:design}. These devices individually show approximately 17 dB of gain, with at least 15 dB gain spanning an octave from 1.5--3.2 GHz, with a nearly identical gain profile for the two devices excluding variations in the standing-wave ripple. As these devices were located roughly 4 cm apart on our wafer (due to our general fabrication strategy of alternating different designs for different purposes), we do not expect them to be exactly identical. These individually measured gain results are sufficient for initial experimental prototyping of the balanced architecture, but could be improved upon further through a dedicated fabrication strategy as noted in the discussion.  

\subsection{Hybrid reference}

\begin{figure}
    \centering
    \includegraphics[width=1\linewidth]{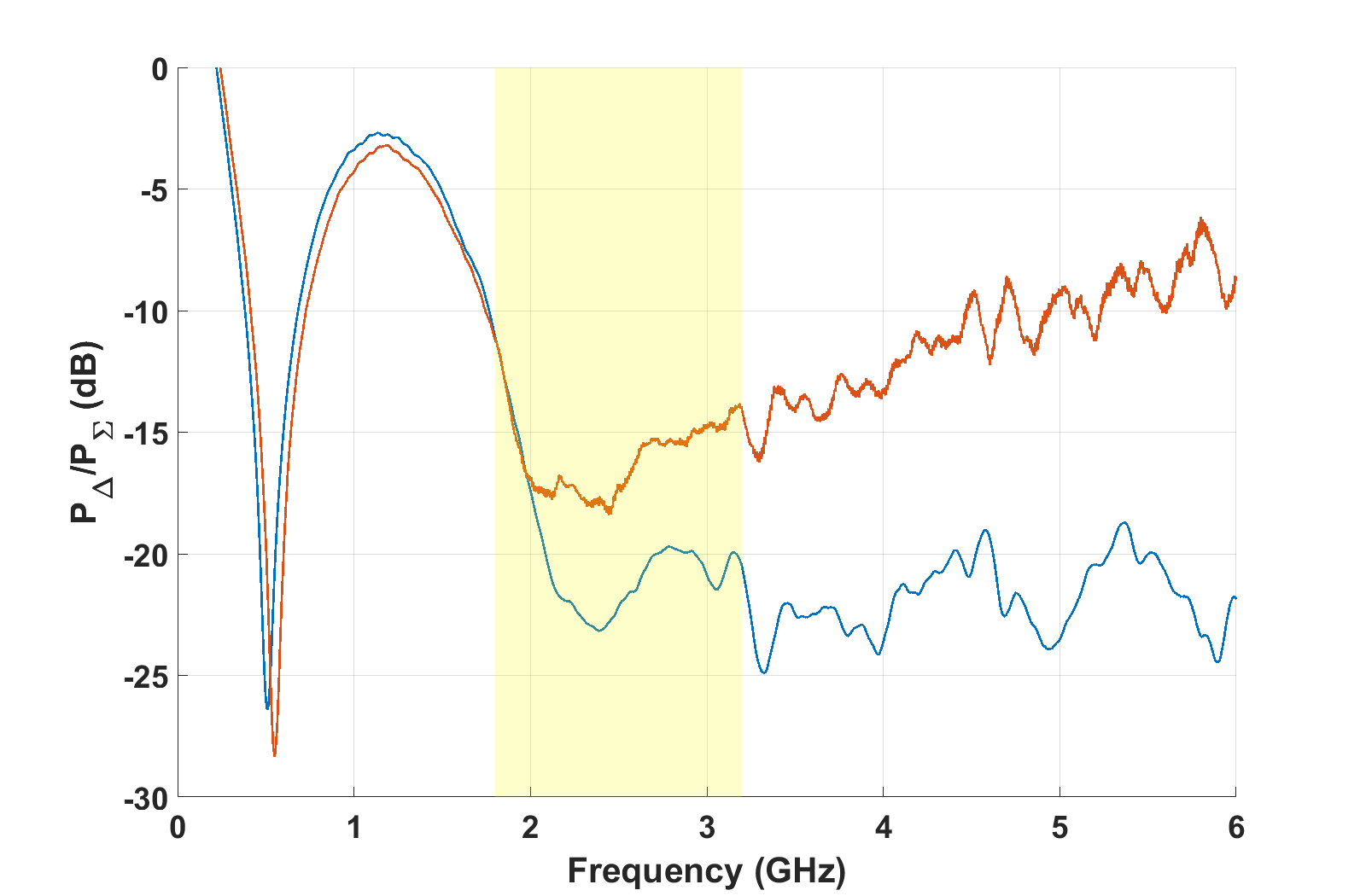}
    \caption{Experimentally measured signal leakage of the balanced TWPA configuration from Figure \ref{fig:BalancedPlot} without the TWPAs (blue) and with the TWPAs (red). The highlighted region represents the potential operating range of our balanced architecture.}
    \label{fig:HybridImbalance}
\end{figure}

For calibration, we first measure the signal/idler isolation provided by our balanced setup without the TWPAs present. This serves as a reference for gain comparison and allows us to measure the imbalance of the hybrids themselves. The results, shown as the blue curve in Figure \ref{fig:HybridImbalance}, provide a metric for the maximum efficacy we can obtain from our experiment. With only the two hybrids connected back-to-back, we immediately see two limitations within the setup. 

At frequencies below 1.7 GHz, which are below the nominal 2-14 GHz operating range of our hybrids, we observe a significant amount of signal leakage through the idler port as expected. In combination with the gain measurements of the TWPAs shown in section~\ref{sec:device_design}, we therefore define the working range for this balanced TWPA from 1.8--3.2 GHz. The lower limit is set by the minimum operating range of the cryogenic 180$^\circ$ hybrids we have available, and the upper limit is set by the maximum frequency at which the TWPA provides gain.

Note that even within the defined operational frequency range, the bare back-to-back hybrid configuration (without TWPAs) still present amplitude/phase imbalance with minimal signal leakage at the level of --20 dB. Therefore, we do not anticipate the image or pump rejection ratio to exceed this level even with a pair of perfectly identical TWPAs in a balanced-TWPA architecture. 

\subsection{Balanced TWPA architecture}

As we introduce the two TWPAs between the hybrids, shown in red in Figure \ref{fig:HybridImbalance}, a small difference in their electrical length introduces a stronger imbalance which scales with frequency. Based on these results, and further corroborated by the inductance measurements detailed in Appendix~\ref{app:inductance}, we can estimate that the phase imbalance in the two TWPAs is below $35^\circ$ at these frequencies (after $\mathcal{O}(10^5)$ degrees of propagation), corresponding to $\epsilon < 0.63$.

\begin{figure}[h]
    \centering
    \includegraphics[width=\linewidth]{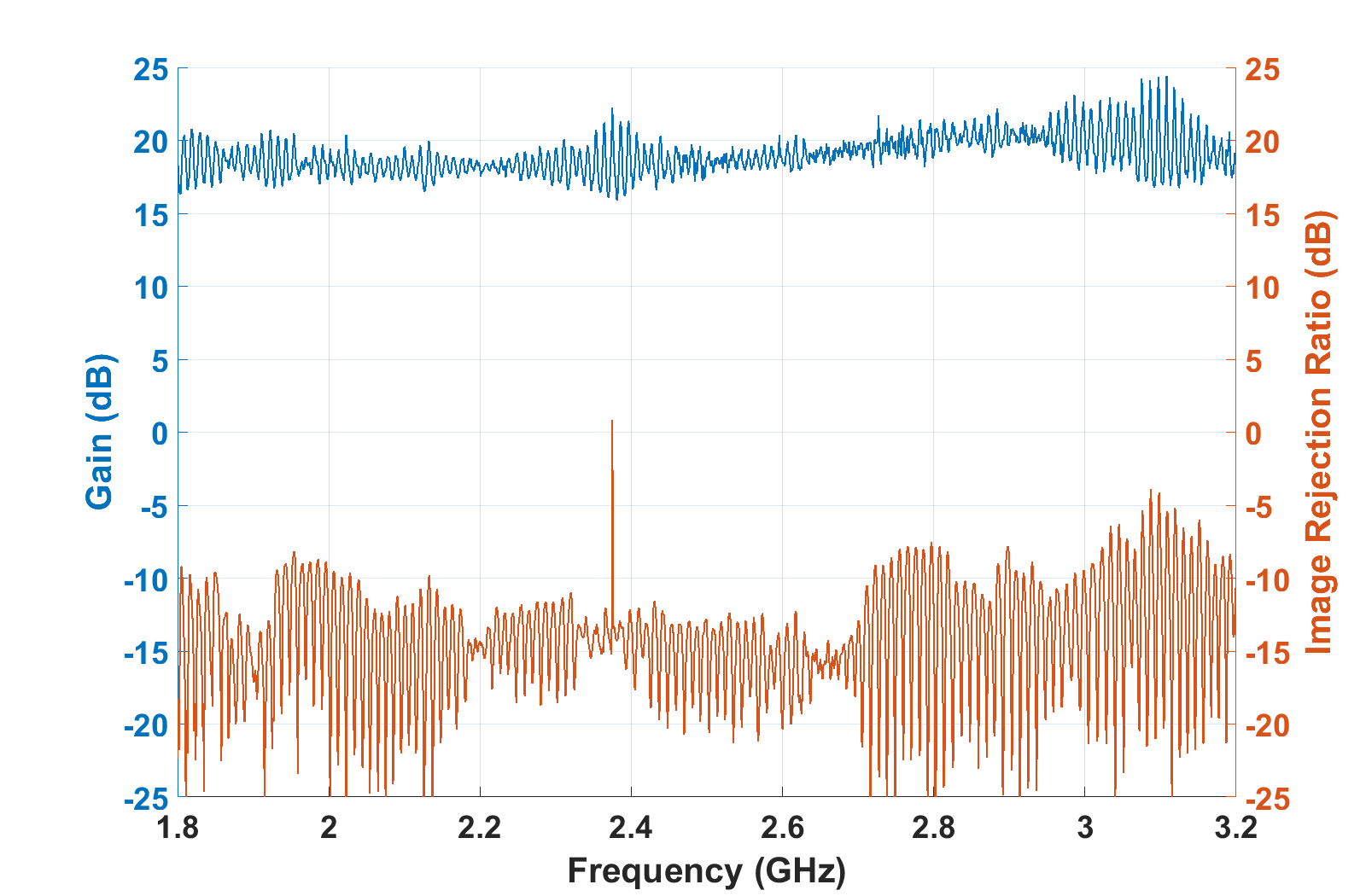}
    \caption{Gain (blue) and signal/idler rejection (red) for a balanced TWPA when optimizing for maximal idler rejection: 4.75 GHz pump and 0.88 mA DC current. Note that since the current is now split between the two TWPAs, the supplied current to the balanced configuration is roughly double the level used by each individual TWPA as shown in Figure~\ref{fig:IndividualTWPAGain}.}
    \label{fig:BalancedZoom}
\end{figure}

\begin{figure}[h]
    \centering
    \includegraphics[width=\linewidth]{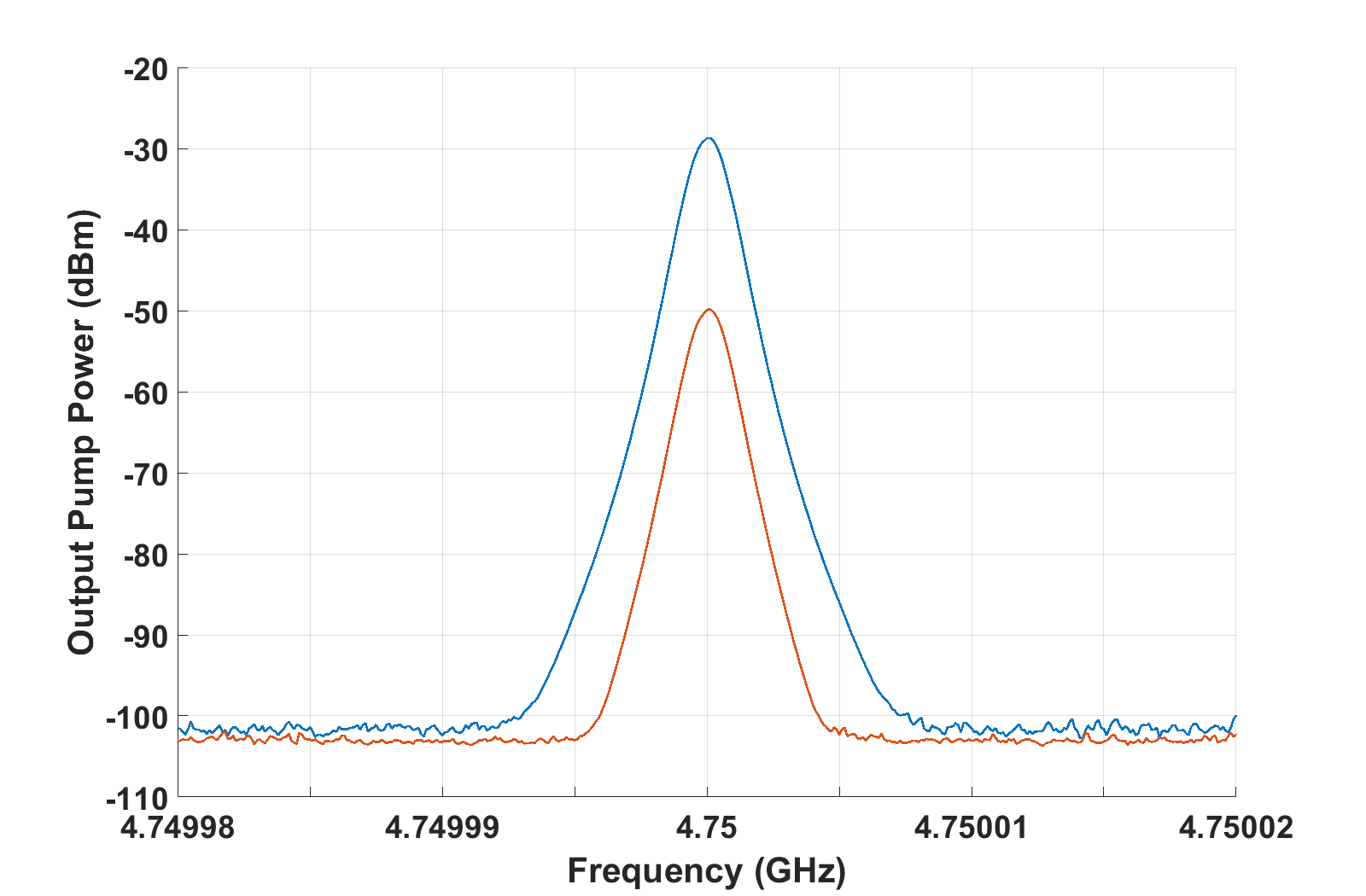}
    \caption{Pump tones measured using a spectrum analyser from the $\Delta$ output port (blue, the pump+idler port) and the $\Sigma$ output port (red, the signal port) showing a pump rejection ratio, $R_p = -21.37$ dB.}
    \label{fig:PumpIsol}
\end{figure}

With the pump supplied to the $\Delta$ input port, the weak signal to the $\Sigma$ input port of the first hybrid, with a single-channel DC current to both TWPAs, we measure the amplitude of the signal, idler, and pump from both the $\Sigma$ and $\Delta$ output ports of the second hybrid (Figures~\ref{fig:BalancedZoom} and \ref{fig:PumpIsol}. These measurements provide us with the gain of the balanced-TWPA ($G_s$), as well as the pump rejection ($R_p$) and image rejection ($R_i$) ratios, as defined in section \ref{sec:imbalanceTheory}. Due to differences in the coaxial cabling from the individual TWPA characterisation performed in section~\ref{sec:device_design}, we slightly retune the pump power and DC bias for the balanced-TWPA configuration in order to re-optimise the gain level. 

The resulting balanced-gain profile is shown in Figure \ref{fig:BalancedZoom}, plotted for the operating range defined (the full gain curve across a broader frequency range is available in Appendix \ref{app:optimisation}). We see that the idler rejection ratio averages at the --15 dB level, closely corresponding to the imbalance measured in the bare back-to-back hybrid configuration shown in Figure~\ref{fig:HybridImbalance}, except at one frequency point at 2.375 GHz: exactly half the pump frequency, $\omega_p/2$, where it is approximately 0 dB. This is an expected result corresponding to the point where the 3WM signal and idler are at the same frequency (in degenerate mode), and thus cannot be spectrally separated by the balanced architecture (see Appendix~\ref{app:degenerate} for details).

With exactly the same configuration, we further achieve a pump rejection ratio in excess of --20 dB, as shown in Figure~\ref{fig:PumpIsol}, between the output $\Sigma$ and $\Delta$ ports. This measurement is taken following room-temperature amplification and calibrated for coaxial cable loss, which we estimate the pump power exiting the balanced TWPA to be roughly 25 dB below these measured values. The leaked pump power to the signal output port is thus well below the saturation power of a typical cryogenic HEMT (high electron mobility transistor) amplifier, post-TWPA-amplification. 

These results demonstrate that the balanced-TWPA architecture is able to simultaneously experimentally provide strong performance across all three key metrics, with
\begin{enumerate}
    \item Signal gain, $G_s \sim 20$ dB,
    \item Pump rejection ratio, $R_p < -20$ dB, and
    \item Image rejection ratio, $R_i \sim -15$ dB.
\end{enumerate}

\section{Discussion}

The results of the previous section demonstrate the experimental realisation of a balanced 3WM TWPA architecture. The balanced TWPA was able to provide 20 dB gain while maintaining good pump and idler rejection levels across an approximately 2:1 bandwidth. These results are encouraging for practical applications that would benefit from separating the amplified signal from the idler and pump tones. 

\subsection{Further improving imbalance}

As described in section~\ref{sec:device_design}, the two TWPAs used in this demonstration were not specifically fabricated or especially screened for balanced operation. The difference in phase velocity between the TWPAs can be significantly improved by optimising the fabrication strategy. 

Gradual variations in the properties of the highly nonlinear superconducting film or the dielectric parameters across the wafer are magnified by this difference, resulting in greater differences between TWPAs with large separations on the wafer. These fabrication variations are well studied, and our earlier work shows their signature is even detectable on the scale of an individual TWPA \cite{Klimovich2024}, where fluctuations in the superconducting film's kinetic inductance result in a measurable frequency shift of the engineered band gap. 

Through the analysis in Appendix \ref{app:inductance}, we estimate that the average normal-state sheet resistance of the superconducting film varied by about 0.05\% between the two TWPAs used in this balanced-TWPA demonstration, resulting in the imbalance shown in Figure \ref{fig:HybridImbalance} and limiting our image rejection ratio to the --15 dB level. These fabrication variations can be substantially mitigated by placing the two TWPAs immediately adjacent to one another on a wafer, ensuring that any variation in superconducting film properties is identical for the two devices \cite{Mantegazzini2023}. 

\subsection{Suppression of reflection-induced gain ripple}
\label{sec:ripple_suppression}

Gain ripples observed in many individual TWPAs arise from weak reflections from impedance mismatches at the two ends of the TWPA and other external components \cite{Chang2025}. A fraction of the forward-propagating signal, pump, and idler is reflected at the output of the device, propagates back towards the input, is reflected again from the TWPA input, and then makes an additional forward pass. The round-trip propagating tones therefore interfere coherently with the directly transmitted tones, and because the accumulated round-trip phase varies with frequency, this interference produces the familiar quasi-periodic modulation of the measured gain. Due to the high gain of the TWPA, this round-trip reflection effect can have a comparable amplitude to the initial input signal, resulting in substantially larger variations in the $S_{21}$ ripple with the amplifier on compared to when the amplifier is off.

Let $\phi_j$ be the one-way propagation phase through TWPA $j$, and let
\begin{equation}
    \bar{\rho} = \rho e^{i 2\phi_j} = g e^{i\phi_j} S_{22} e^{i \phi_j} S_{11} 
\end{equation}
denote the effective common round-trip coefficient associated with an individual TWPA, neglecting loss and noting that the gain is applied only to the wave that is co-directional with the pump. The field leaving arm $j$ is then
\begin{align}
    B_j
    &=
    \frac{gA_s}{\sqrt{2}}
    \left[
        e^{i\phi_j}
        +\rho e^{i3\phi_j}
        +\rho^2e^{i5\phi_j}
        +\cdots
    \right] \nonumber\\
    &=
    \frac{gA_s}{\sqrt{2}}
    \frac{e^{i\phi_j}}
         {1-\rho e^{i2\phi_j}},
    \label{eq:Bj_ripple}
\end{align}
for $|\rho|<1$. The first term has traversed the TWPA once, while the first round-trip contribution has traversed it three times: forward, backward, and forward again. Each subsequent round trip adds two further traversals. For a single TWPA, the factor $(1-\rho e^{i2\phi_j})^{-1}$ therefore produces the familiar standing-wave modulation of the gain, which has a divergent behaviour as $g$ increases and $|\rho|$ approaches 1. For the balanced amplifier, the output hybrid coherently combines the fields of the TWPAs. The signal at the sum port is therefore
\begin{equation}
    A_{\Sigma,\mathrm{out}}
    =
    \frac{B_1+B_2}{\sqrt{2}}.
\end{equation}
\begin{equation}
    A_{\Sigma,\text{out}}
    =
    \frac{gA_s}{2}
    \left[
        \frac{e^{i\phi_1}}
             {1-\rho e^{i2\phi_1}}
        +
        \frac{e^{i\phi_2}}
             {1-\rho e^{i2\phi_2}}
    \right].
    \label{eq:Asigma_split}
\end{equation}
Substituting Eq.~\eqref{eq:Bj_ripple} gives
\begin{align}
    A_{\Sigma,\mathrm{out}}
    =
    gA_s
    \sum_{n=0}^{\infty}
    \rho^n
    e^{i(2n+1)\bar{\phi}}
    \cos\left[
        \frac{(2n+1)\delta}{2}
    \right]
    \label{eq:ripple_balanced_series}
\end{align}
where 
\begin{equation}
    \bar{\phi}=\frac{\phi_1+\phi_2}{2},
    \qquad
    \delta=\phi_1-\phi_2.
\end{equation}
For convenience, we can extract the $n = 0$ term, corresponding to the single-pass gain, resulting in
\begin{equation}
    A_{\Sigma}
    =
    gA_s e^{i\bar{\phi}}
    \cos\left(\frac{\delta}{2}\right)
    \sum_{n=0}^{\infty}
    \left(\rho e^{i2\bar{\phi}} \right)^n
    \frac{
        \cos\left[(2n+1)\delta/2\right]
    }{
        \cos\left(\delta/2\right)
    },
    \label{eq:HRT_series}
\end{equation}
where the $n$th term corresponds to a field that has traversed the TWPA $2n+1$ times and consequently accumulates an arm-to-arm phase difference $(2n+1)\delta$. 

The directly transmitted signal scales as $\cos(\delta/2)$, whereas the first reflected contribution scales as $\cos(3\delta/2)$, the second as $\cos(5\delta/2)$, and so on. Relative to the directly transmitted carrier, the $n$th ripple contribution is reduced from its independent TWPA's counterpart by
\begin{equation}
    \mathcal{R}_n
    =
    \left|
    \frac{
        \cos\left[(2n+1)\delta/2\right]
    }{
        \cos\left(\delta/2\right)
    }
    \right|.
    \label{eq:ripple_suppression_factor}
\end{equation}
Thus, multiply reflected fields decorrelate between the two arms in a balanced architecture far more rapidly than the single-pass signal in an individual TWPA. This means that even small imbalances in the relative phase of the two arms, $\delta$, would result in suppression of the gain ripple. This explains the results of Figure \ref{fig:BalancedZoom}, where we see a reduction in gain ripple compared to the individual TWPA case in the 2.4--3 GHz regime. 

This occurs primarily in the frequency range at which the imbalance of the two TWPAs starts to increase, as shown in Figure \ref{fig:HybridImbalance}. The reflected field removed from the $\Sigma$ port is redirected by the hybrid to the $\Delta$ port with amplitude 
\begin{align}
    A_{\Delta}
    &=
    \frac{gA_s}{2}
    \left[
        \frac{e^{i\phi_1}}
             {1-\rho e^{i2\phi_1}}
        -
        \frac{e^{i\phi_2}}
             {1-\rho e^{i2\phi_2}}
    \right] \nonumber\\
    &=
    i gA_s e^{i\bar{\phi}}
    \sum_{n=0}^{\infty}
    \rho^n e^{i2n\bar{\phi}}
    \sin\left[
        \frac{(2n+1)\delta}{2}
    \right].
    \label{eq:Adelta_ripple}
\end{align}
This is an interesting finding, as it means that small levels of imbalance between the two TWPAs could potentially be beneficial. Depending on the particular application, there is a possible trade-off between optimising for very high levels of idler rejection or, alternatively, leaving small imbalances and benefiting from the slightly reduced gain ripple. 

\section{Conclusion}

We have experimentally demonstrated a balanced three-wave-mixing travelling-wave parametric amplifier in which two nominally identical kinetic-inductance TWPAs are combined using a pair of $180^{\circ}$ hybrid couplers. By supplying the signal and pump through the $\Sigma_\text{in}$ and $\Delta_\text{in}$ ports, respectively, the relative phases of the fields in the two amplifier arms result in the amplified signal recombining at the $\Sigma_\text{out}$ port, while the pump and the unwanted idler are redirected to the $\Delta_\text{out}$ port. This provides a practical means of separating the signal from the idler and pump without sacrificing the original gain-bandwidth performance of TWPAs. 

In our initial implementation, the balanced-TWPA provides approximately $20\,\mathrm{dB}$ of signal gain over nearly an octave bandwidth, while simultaneously achieving a pump rejection ratio below $-20\,\mathrm{dB}$ and an average image rejection ratio of approximately $-15\,\mathrm{dB}$ across the defined operating band. These results establish the balanced TWPA as a practical route towards broadband, low-noise parametric amplification with intrinsic separation of the amplified signal from the unwanted pump and idler, alongside potential improvements in the gain ripple. Such a capability is particularly attractive for applications involving broadband or a priori unknown input spectra, including astronomical heterodyne receivers and highly multiplexed cryogenic detector readout systems.

\section{Acknowledgements}

For the purpose of Open Access, the author has applied a CC BY public copyright licence to any Author Accepted Manuscript version arising from this submission. 

\section*{Funding}

We thank the Leverhulme Trust Visiting Professorship (VP1-2024-022), UKRI International Science Partnerships Fund (ISPF) Quantum for Science Bilateral Visit Grant (Project No. ST/Y004973/1), STFC Astronomy Observation and Theory Small Award (Project No. ST/Y002210/1), MERAC Foundation, and European Research Council (ERC) Consolidator Grant (SPARTAN-FX, 101169947) \& Starter Grant (SPA4AstroQIT, 803862) for funding of this work. Views and opinions expressed are however those of the author(s) only and do not necessarily reflect those of the European Union or the European Research Council. Neither the European Union nor the granting authority can be held responsible for them.

\section*{Appendix}

\subsection{TWPA Design Parameters}
\label{app:design}

Table~\ref{tab:parameters} shows the design parameters of the two TWPAs used in this work, including material properties of the NbTiN conductor and ground (normal state resistivity $\rho$, and critical temperature $T_c$); central conductor, dielectric, and ground plane thicknesses ($t_{ms}$, $t_\epsilon$, $t_{gnd}$); central conductor width ($w$); and capacitive stub spacing ($d$), average length ($l_0$), modulation amplitude and period ($l_a$, $\lambda_p$). See Ref.~\cite{Klimovich2024} for other geometric details.

\begin{table*}[t]
    \centering
    \caption{}
    \begin{tabularx}{\textwidth}{CCCCCCCCCC}
        \toprule
        \multicolumn{2}{c}{NbTiN Parameters} & \multicolumn{3}{c}{Film Thickness (nm)} & \multicolumn{5}{c}{Central Conductor Geometry ($\mu$m)} \\
        \cmidrule(r){1-2} \cmidrule(r){3-5} \cmidrule(r){6-10}
        $\lambda $ & $T_c$ & $t_{ms}$ & $t_\epsilon$ & $t_{gnd}$  & $w$ & $d$ & $l_0$ & $l_a$ & $\lambda_p$ \\
        \addlinespace
        370 nm & 14.5 K & 35 & 190 & 200 & 0.25 & 2 & 14 & 2 & 87.5\\
        \bottomrule
    \end{tabularx}
    \label{tab:parameters}
\end{table*}

\subsection{Fabrication constraints}
\label{app:inductance}

\begin{figure}[h]
    \centering
    \includegraphics[width=\linewidth]{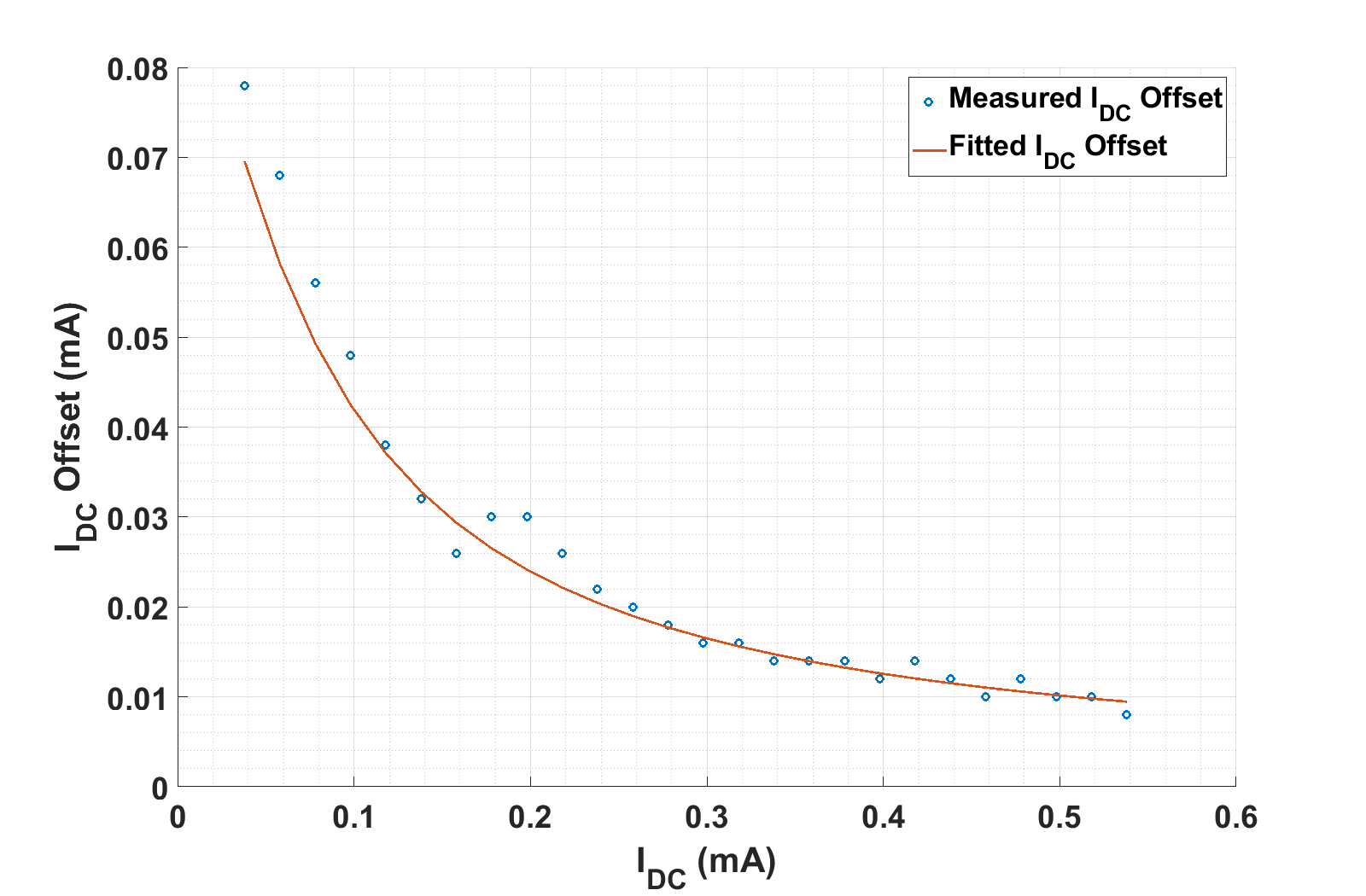}
    \caption{The excess DC current that is supplied to one TWPA to match the phase delay of the second, plotted as a function of the base level of DC current supplied to both. The fitted line shows the expected offset given a kinetic inductance ratio of $L_{k,1} = 0.9995 L_{k,2}$ between the two TWPAs.}
    \label{fig:OffsetChart}
\end{figure}

The two TWPAs used for this preliminary demonstration were taken from part of the wafer several centimetres apart, amplifying these differences in their superconducting film inductances, and thereby phase velocities, $v_\text{ph} = 1/\sqrt{LC}$. We can attempt to measure the difference in the phase velocity of the two TWPAs' lines, by utilising the nonlinear response of the kinetic inductance to applied current:
\begin{equation}
    L_k = L_0 \left(1 + \frac{I^2}{I_*^2}\right).
\end{equation}
By supplying various magnitudes of DC current to each TWPA and measuring the phase delay of the signals that propagate through them, we can calibrate the excess DC current we need to supply for one device to match the phase velocity of the second. The results are depicted in Figure~\ref{fig:OffsetChart}, showing the excess current that needs to be supplied to one device to match the phase velocity (hence the kinetic inductance) of the second. This translates to about 0.05\% difference in electrical length between the two TWPAs used in this experiment.

The unpumped kinetic inductance is a function of the normal state sheet resistance, $R_n$, and the superconducting gap, $\Delta_0$. 
\begin{equation}
    L_0 = \frac{\hbar R_n}{\pi\Delta_0}
\end{equation}
Importantly, this is a constraint on the variation of the kinetic inductance between the two TWPAs and not the precision to which the inductance must be controlled within each fabrication run. In principle, the superconducting film parameters can have a large gradient across the wafer as long as this gradient is consistent between the two TWPAs. Given the efficacy of the results presented here, we believe this can be achieved by fabricating the two devices nearer to each other on the wafer.

\subsection{Broadband gain curve}
\label{app:optimisation}

\begin{figure}[h]
    \centering
    \includegraphics[width=\linewidth]{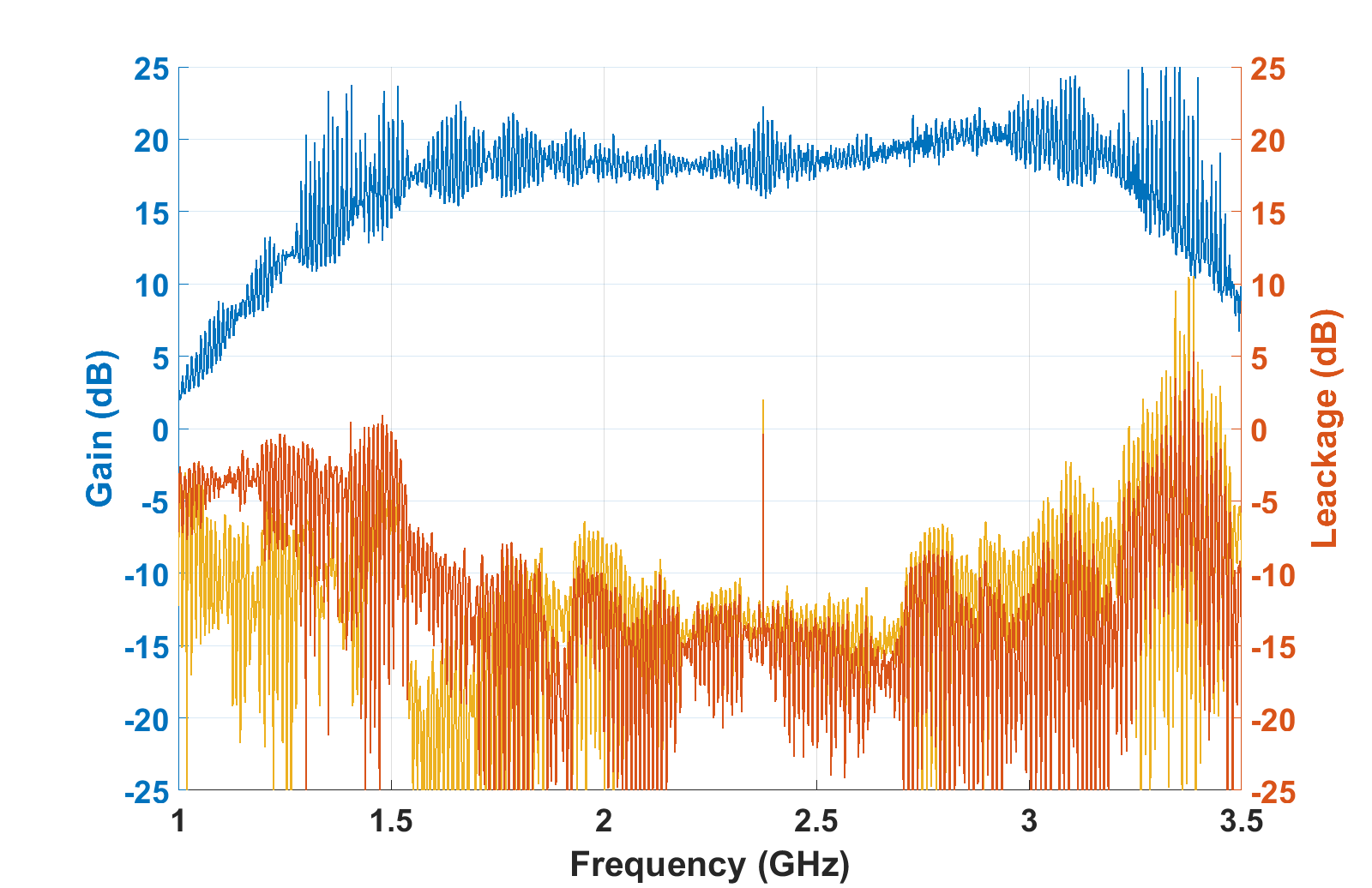}
    \caption{Gain (blue) and signal/idler rejection (red) for a balanced TWPA when optimizing for maximal idler rejection: 4.75 GHz, 6.5 dBm pump, 0.88 mA DC current.}
    \label{fig:BalancedBroad}
\end{figure}

Figure \ref{fig:BalancedBroad} shows the gain profile and leakage of the balanced TWPA scheme operated in a similar biasing condition to the gain measured for each individual TWPA in Figure~\ref{fig:IndividualTWPAGain}, plotted on the same frequency scale. As expected, the leakage level substantially decreases outside of the operating range of the hybrids identified in the reference measurements in Figure~\ref{fig:HybridImbalance}.

The leakage levels near the sides of the gain curve are a product of the minimum working regime of our $180^\circ$ hybrids, as shown in Figure~\ref{fig:HybridImbalance}. This affects the balanced tone propagating at lower frequencies, i.e., the signal leakage in red, and the idler image rejection shown in yellow. Similarly, the degraded $S_{11}$ at these frequencies results in stronger reflections and thereby a larger gain ripple compared to the individual TWPA characterisation without the hybrids.

\subsection{Behaviour at the degenerate three-wave-mixing point}
\label{app:degenerate}

The signal--idler separation described in Sec.~\ref{sec:theory} assumes non-degenerate three-wave mixing, for which the signal and idler occupy distinct modes at frequencies $\omega_s$ and $\omega_i=\omega_p-\omega_s$. At the degenerate point,
\begin{equation}
    \omega_s=\omega_i\equiv\omega_d=\frac{\omega_p}{2},
\end{equation}
this distinction is lost. The signal and idler are then two coherent contributions to the same mode, and the assumption used in Sec.~\ref{sec:theory} that the amplified signal may be treated independently from the generated idler is no longer valid.

For a perfectly phase-matched degenerate 3WM process, the coupled-mode equation can be written in the form
\begin{equation}
    \frac{dA}{dz}=i\kappa A_p A^*,
    \label{eq:degenerate_cme}
\end{equation}
where $A$ is the complex amplitude of the mode at $\omega_d$. Its solution through either TWPA can be expressed as
\begin{equation}
    A_{\mathrm{out}}=\mu A_{\mathrm{in}}+\nu A_{\mathrm{in}}^*,
    \label{eq:degenerate_bogoliubov}
\end{equation}
with $|\mu|^2-|\nu|^2=1$. The coefficient $\nu$ contains the pump phase and therefore changes sign when the pump phase is shifted by $\pi$. In the balanced configuration, the pumps in the two arms are in anti-phase, so that, for otherwise identical TWPAs,
\begin{equation}
    \nu_2=-\nu_1\equiv-\nu,
\end{equation}
while $\mu$ is the same in both arms.

At $\omega_d$, an incident field applied to the $\Sigma$ port of the first hybrid is divided equally and in phase between the two TWPAs,
\begin{equation}
    A_{1,\mathrm{in}}=A_{2,\mathrm{in}}=\frac{A_\Sigma}{\sqrt{2}}.
\end{equation}
The fields immediately before the output hybrid are therefore
\begin{align}
    A_{1,\mathrm{out}}
        &=\frac{1}{\sqrt{2}}\left(\mu A_\Sigma+\nu A_\Sigma^*\right),\\
    A_{2,\mathrm{out}}
        &=\frac{1}{\sqrt{2}}\left(\mu A_\Sigma-\nu A_\Sigma^*\right).
\end{align}
Recombination in the second $180^\circ$ hybrid gives
\begin{align}
    A_{\Sigma,\mathrm{out}}
        &=\frac{A_{1,\mathrm{out}}+A_{2,\mathrm{out}}}{\sqrt{2}}
          =\mu A_\Sigma,\\
    A_{\Delta,\mathrm{out}}
        &=\frac{A_{1,\mathrm{out}}-A_{2,\mathrm{out}}}{\sqrt{2}}
          =\nu A_\Sigma^*.
    \label{eq:degenerate_outputs}
\end{align}
Thus, even for a perfectly balanced device, both output ports contain a coherent field at the same frequency $\omega_p/2$. The $\Sigma$ output contains the component proportional to the incident field, whereas the $\Delta$ output contains its phase-conjugated parametric component. These two terms would be identified as signal and idler away from degeneracy, but at $\omega_s=\omega_i$ they are not distinct spectral modes and therefore cannot be separated in the sense used for the non-degenerate case.

Consequently, there is no isolated output port at the degenerate frequency, $\omega_d$. Instead, the output powers are proportional to $|\mu|^2|A_\Sigma|^2$ and $|\nu|^2|A_\Sigma|^2$ at the $\Sigma$ and $\Delta$ ports, respectively. In the high-gain limit $|\mu|^2\simeq|\nu|^2$, so comparable power is expected at both outputs. Using the definition of image rejection, $R_i$, from Sec.~\ref{sec:theory}, this results in an expected $R_i = 1$ (0 dB) at $\omega_d = \omega_p/2$, exactly as is seen in our experimental results in Figure~\ref{fig:BalancedBroad}.

\bibliographystyle{iopart-num}

\bibliography{balanced_twpa_references}

\end{document}